\documentclass{PoS}
\usepackage[square,sort,comma,numbers]{natbib}
\usepackage{graphicx,epstopdf}

\newcommand{\hi}{H~{\sc i}}
\newcommand{\kms}{km\,s$^{-1}$}

\title{The MeerKAT Absorption Line Survey (MALS)}

\ShortTitle{MALS}

\author{
\speaker{N. Gupta}$^1$, 
R. Srianand$^1$,  
W. Baan$^2$,
A. Baker$^3$,
R. Beswick$^4$,
S. Bhatnagar$^5$,
D. Bhattacharya$^1$,
A. Bosma$^6$,
C. Carilli$^5$,
M. Cluver$^7$,
F. Combes$^8$,
C. Cress$^9$,
R. Dutta$^1$,
J. Fynbo$^{10}$,
G. Heald$^{11}$,
M. Hilton$^{12}$,
T. Hussain$^1$,
M. Jarvis$^{7,13}$
G. Jozsa$^{14}$,
P. Kamphuis$^{15}$,
A. Kembhavi$^1$,
J. Kerp$^{16}$,
H.-R. Kl\"ockner$^{17}$,
J. Krogager$^{18}$,
V. Kulkarni$^{19}$,
C. Ledoux$^{20}$,
A. Mahabal$^{21}$,
T. Mauch$^{14}$,
K. Moodley$^{12}$,
E. Momjian$^5$,
R. Morganti$^2$,
P. Noterdaeme$^{18}$,
T. Oosterloo$^2$,
P. Petitjean$^{18}$,
A. Schr\"oder$^{22}$,
P. Serra$^{23}$,
J. Sievers$^{12}$,
K. Spekkens$^{24}$,
P. V\"{a}is\"{a}nen$^{22}$,
T. van der Hulst$^{25}$,
M. Vivek$^{26}$,
J. Wang$^{11}$,
O.I. Wong$^{27}$
and A.R. Zungu$^{12}$

\\
$^1$ Inter-University Centre for Astronomy and Astrophysics, India; \\(see Appendix A for author affiliations.)  

E-mail: \email{ngupta@iucaa.in}
       }

\abstract{
Deep galaxy surveys have revealed that the global star formation rate (SFR) density in the Universe peaks 
at 1$\le z \le$2 and sharply declines towards $z = 0$. But a clear picture of the underlying processes, in particular 
the evolution of cold atomic ($\sim$100\,K) and molecular gas phases, that drive such a strong evolution is yet to emerge. 
MALS is designed to use MeerKAT's L- and UHF-band receivers to carry out the most sensitive (N(\hi)$>$10$^{19}$\,cm$^{-2}$) 
dust-unbiased search of {\it intervening} \hi\ 21-cm and OH 18-cm absorption lines at $0<z<2$.  
This will provide reliable measurements of the evolution of cold atomic and molecular gas cross-sections of galaxies, 
and unravel the processes driving the steep evolution in the SFR density. 
The large sample of \hi\ and OH absorbers obtained from the survey will (i) lead to tightest constraints on the fundamental 
constants of physics, and (ii) be ideally suited to probe the evolution of magnetic fields in disks 
of galaxies via Zeeman Splitting or Rotation Measure synthesis.  
The survey will also provide an unbiased census of \hi\ and OH absorbers, i.e. cold gas {\it associated} with powerful 
AGNs ($>$10$^{24}$\,W\,Hz$^{-1}$) at $0<z<2$, and will simultaneously deliver a blind \hi\ and OH emission 
line survey, and radio continuum survey. 
Here, we describe the MALS survey design, observing plan and the science issues to be addressed under various science 
themes.   
}

\FullConference{MeerKAT Science: On the Pathway to the SKA,\\
	25-27 May, 2016,\\
		Stellenbosch, South Africa}

\begin{document}
%%%%%%%%%%%%%%%%%%%%%%%%%%%%%%%%%%%%%%%%%%%%%%%%%%%%%%

%%%%%%%%%%%%%%%%%%%%%%%%%%%%%%%%%%%%%%%%%%%%%%%%%%%%%%
\section{Background}
%%%%%%%%%%%%%%%%%%%%%%%%%%%%%%%%%%%%%%%%%%%%%%%%%%%%%%

In recent years, tremendous efforts have been dedicated to establish the evolution of the global 
comoving SFR density up to $z\sim8$. The observations show that there is a peak in the comoving SFR 
density between $1<z<3$, followed by an order of magnitude decrease towards $z\sim$0 over the last 10\,Gyrs,  
i.e. ~70\% of the age of the Universe \citep[][]{Bouwens14}. But a clear picture of the underlying processes 
that drive such a strong evolution is yet to emerge.

Since stars form in molecular clouds, the observations of the molecular gas (H$_2$), which is the basic 
fuel for star formation, and the atomic gas (\hi), which is the dominant gas component of galactic discs, 
can be used to understand the evolution of SFR density.  The atomic gas component of the interstellar 
medium (ISM) is easily observable through the \hi\ 21-cm emission line. But due to technical 
limitations the \hi\ 21-cm emission line studies at present are mostly confined to the local Universe. 
Through the observations of \hi\ 21-cm emission line in the local Universe and the large samples of damped 
Ly$\alpha$ absorbers (DLAs\footnote{\hi\ column density, N(\hi)$\ge$2$\times$10$^{20}$\,cm$^{-2}$.}) 
at high-$z$, the \hi\ mass density ($\Omega_{\rm {HI}}$) of the Universe is reasonably 
well-constrained at $z<0.2$ and $2<z<4$, respectively. They reveal a very modest (factor $\sim$2) decrease 
in $\Omega_{\rm {HI}}$ compared to an order of magnitude decrease in the SFR density over the same redshift range. 
This implies that the processes leading to the conversion of atomic gas to molecular gas and eventually to 
stars need to be understood directly via observations of cold atomic and molecular gas, rather than the 
evolution of $\Omega_{\rm {HI}}$.

It is well known that \hi\ 21-cm absorption is an excellent tracer of the cold neutral medium (CNM; T$\sim$100 K) 
in the Galaxy. Similarly, the OH radical is one of the common constituents of diffuse and dense molecular 
clouds, and can be observed through the two  main lines at 1665 MHz and 1667 MHz, and the two satellite lines at 
1612 MHz and 1720 MHz. As the volume filling factor of ISM phases and the observables associated with 
the \hi\ and OH absorption lines depend sensitively on in situ star formation through various stellar 
feedback processes, these lines can be used to probe the physical conditions in the ISM and understand 
the cosmic evolution of the SFR density \citep[][]{Liszt96, Heiles03, Wolfire03, Ledoux15}.
In addition, the observed frequencies of these lines can be used to place the most stringent constraints 
on the variation of fundamental constants of physics \citep[][]{Kanekar05, Rahmani12} and cosmic 
acceleration \citep[][]{Darling12, Kloeckner15}.

%%%%
Despite the scientific potential of intervening \hi\ 21-cm and OH 18-cm lines, and the tremendous efforts 
from the community over last three decades, mainly due to technical limitations, only $\sim$50 
intervening \hi\ 21-cm \citep[][]{Gupta12, Kanekar14} and 3 intervening molecular absorbers \citep[][]{Combes08} 
at radio wavelengths are known. The small number of detections has hampered attempts to systematically 
map the cold gas evolution with any statistical significance \citep[Fig. 13 of][]{Gupta12}.

%%%%
{\it The availability of various SKA pathfinders in RFI-free environments offers an unique opportunity 
to overcome these limitations.}

%%%%
In 2010, MALS was identified as one of the ten large surveys to be carried out with MeerKAT.  
The main objective of the survey is to use MeerKAT's L- and UHF-band receivers to carry out the most sensitive 
(N(\hi)$>$10$^{19}$\,cm$^{-2}$) dust-unbiased search of {\it intervening} \hi\ 21-cm and OH 18-cm absorption lines at $0<z<2$.  
The survey is expected to detect $\sim$200 intervening 21-cm absorbers. 
This will provide reliable measurements of the evolution of cold atomic and molecular gas cross-sections of galaxies, 
and unravel the processes driving the steep evolution in the SFR density. 
Due to the large field-of-view and excellent sensitivity of MeerKAT, the survey is also expected to detect 
$\sim$500 21-cm absorbers {\it associated} with AGNs. 
%%% 
These {\it associated} and {\it intervening} absorbers detected from the survey will enable discovery through 
a wide range of science themes that are described in Section~2. 
The survey design is presented in Section~3.   
%%%
A high-level plan for data products, observing plan and data releases is provided in Section~4.  
The relationship with other SKA pathfinder surveys and key science is presented in Section~5.    
   
%%%%%%%%%%%%%%%%%%%%%%%%%%%%%%%%%%%%%%%%%%%%%%%%%%%%%%
\section{MALS - Key Science Themes}
\label{sec:themes}
%%%%%%%%%%%%%%%%%%%%%%%%%%%%%%%%%%%%%%%%%%%%%%%%%%%%%%
The key science themes based on the survey design (see Section~3; Table~\ref{survey}) are as follows: 

%%%%
\begin{enumerate}
%%%%

%%%%%
\item{\bf Evolution of cold gas in galaxies and relationship with SFR density:}
MALS will be sensitive to detect \hi\ 21-cm absorption and constrain the number per unit redshift range of 
intervening \hi\ 21-cm absorbers, $n_{21}$, from sub-DLAs to DLAs, i.e. the CNM fraction in a wide range of 
column densities tracing the inner disks of galaxies, as well as the extended disks, halos, intra-group gas 
and the circumgalactic medium (CGM). 
It will measure the evolution of cold atomic and molecular gas cross-section of galaxies in a luminosity- and 
dust-unbiased way and constrain the ISM physics at pc scales over the entire redshift range $0<z<2$.
%%%
Through MALS, the detection of $\sim$200 intervening 21-cm absorbers is expected. With these it will 
constrain the evolution of $n_{21}$($z$), split in 3 redshift bins over $0<z<1.5$, with $\sim$20\% accuracy.
If $n_{21}$($z$) scales with $\Omega_{\rm {HI}}$($z$), the 20\% accuracy will allow us to detect the evolution in 
$n_{21}$($z$), and hence the CNM cross-section of galaxies, at the 3$\sigma$ level (for example, see \citep[][]{Allison16}). 
Through OH lines, the survey 
will be sensitive to the molecular gas cross-section of galaxies, and will provide the first estimate of the number 
per unit redshift range of OH absorbers, $n_{OH}$($z$).  The $n_{21}$($z$) and $n_{OH}$($z$) from the survey 
will be key observables to understand the process of condensation of warm gas into cold gas and eventually stars.  
%MALS will achieve all this in a dust- unbiased way and 
%will probe all i.e. diffuse, translucent and dense phases of ISM. The latter two have been generally missed out 
%in optical/ultraviolet surveys, including SDSS, which are wellknown to be biased against dust extinction 
%\citep[][]{Noterdaeme10co}.
%%%%%%%%%%%%%%%%%%%%%%%%%%%%%%%%
\begin{table}
\centering
%\begin{minipage}{120mm}
\caption{Summary of various upcoming \hi\ 21-cm absorption line surveys}
\begin{tabular}{cccccccc}
\hline
{\large \strut}  Survey  & Frequency   &  Redshift   & Time   &Spectral          & Sky      & Total  & No. of     \\
                         & coverage    &  range      &  per   &rms per           & coverage & time   & sight lines$^\ddag$\\
                         &             &             &pointing&$\sim$5\,\kms\    &          &        &            \\
                         &   (MHz)     & (\hi\ 21-cm)& (hrs)  &(mJy/b)$^\dag$& (deg$^2$) &  (hrs) &            \\
\hline           
{\large \strut}  Apertif       & 1130 - 1430 & 0   - 0.26 & 12     & 1.3    & 4000 & 6000 & 25000              \\ 
                 SHARP         &             &            &        &        &      &      &       ($>$30\,mJy) \\ 
                 ASKAP         &  700 - 1000 & 0.5 - 1.02 &  2     & 3.8    &25000 & 1600 & 65000              \\ 
                 FLASH         &             &            &        &        &      &      &       ($>$90\,mJy) \\ 
                 ASKAP         & 1130 - 1430 & 0   - 0.26 &  8     & 1.6    &30000 & 8000 &132000              \\ 
                 WALLABY       &             &            &        &        &      &      &       ($>$40\,mJy) \\ 
                 MALS          &  900 - 1670 & 0   - 0.57 &  1     & 0.5    & 1000 &  691 & 12000              \\ 
                 L-band        &             &            &        &        &      &      &       ($>$15\,mJy) \\ 
                 MALS          &  580 - 1015 & 0.4 - 1.44 &  2     & 0.6    &  700 &  746 & 12000              \\ 
                 UHF-band      &             &            &        &        &      &      &       ($>$15\,mJy) \\ 
\hline
\end{tabular}
\flushleft{{\small $^\dag$ Estimated at the center of the band; $^\ddag$ See text for details. \\
} } 
%\flushleft{Column 1: slit orientation; column 2: date of observations; column 3: exposure time in min; 
%column 4 and 5: air mass and grism respectively.}
\label{abs}
\end{table}
%%%%%%%%%%%%%%%%%%%%%%%%%%%%%%%%
%%

In Table~\ref{abs}, we compare MALS with other upcoming \hi\ imaging surveys with footprint and survey depth 
large enough to deliver a substantial number of 21-cm absorbers.   
The Westerbork Synthesis Radio Telescope (WSRT) shallow survey will search for \hi\ emission 
line at $\delta$$>$+30$^\circ$ with Apertif, a phased-array feed for WSRT. Simultaneously, it will deliver a 21-cm 
absorption line survey called SHARP (PI: Morganti).  The Australian SKA Pathfinder (ASKAP) will have a dedicated 
21-cm absorption line survey at $0.5<z<1$ called the First Large Absorption Survey in \hi\ (FLASH; PI: Sadler). 
WALLABY (PIs: Koribalski and Staveley-Smith) is the ASKAP \hi\ emission all-sky survey at $\delta$$<$30$^\circ$ 
that will simultaneously deliver a 21-cm absorption line survey at $z<0.26$.   

From Table~\ref{abs} it is clear that amongst various upcoming major \hi\ 21-cm absorption line surveys, MALS is 
well distinguishable. It is the only survey that will uniformly cover $0<z<1.5$, the most important redshift range 
as far as the evolution of SFR and AGN activity are concerned (cf. columns 2-3). It also provides excellent spectral 
line sensitivity in relatively much less observing time per pointing (cf. columns 4-5). The last column provides the 
number of sources or sight lines within the survey area towards which \hi\ 21-cm absorption from a 100\,K gas with 
the DLA column density would be detectable. The flux density cut-off corresponding to this optical depth sensitivity 
is also provided. Note that the last column does not take into account the redshift distribution of radio sources and 
any variation in gas properties with redshift or radio source type, but provides a useful metric to compare the overall 
effectiveness of these surveys with different characteristics. Fig.~\ref{dist} shows the distribution of
redshift path for intervening absorption for the same optical depth sensitivity. Here, the analysis takes into 
account the redshift and spatial distribution of radio sources over the full survey area. For MALS, we have 
considered the survey area corresponding to 20\% sensitivity of the maximum of the primary beam. In summary, it 
is evident that except for WALLABY for which a large amount of telescope time in a narrow redshift range has been 
allocated to an \hi\ 21-cm emission line survey, the MALS is comparable to other surveys at $z>0.26$. 
In particular, both in the L- and UHF- bands it covers more than 50\% of the redshift ranges inaccessible to any of 
the other surveys.  Notably, the redshift ranges of $0.3<z<0.5$ and $1<z<1.5$ are only accessible to MALS.

%%
%%
%%%%%%%%%%%%%%%%%%%%%%%%%%%%
\begin{figure}
\hbox{
\includegraphics[width=70mm, angle=0]{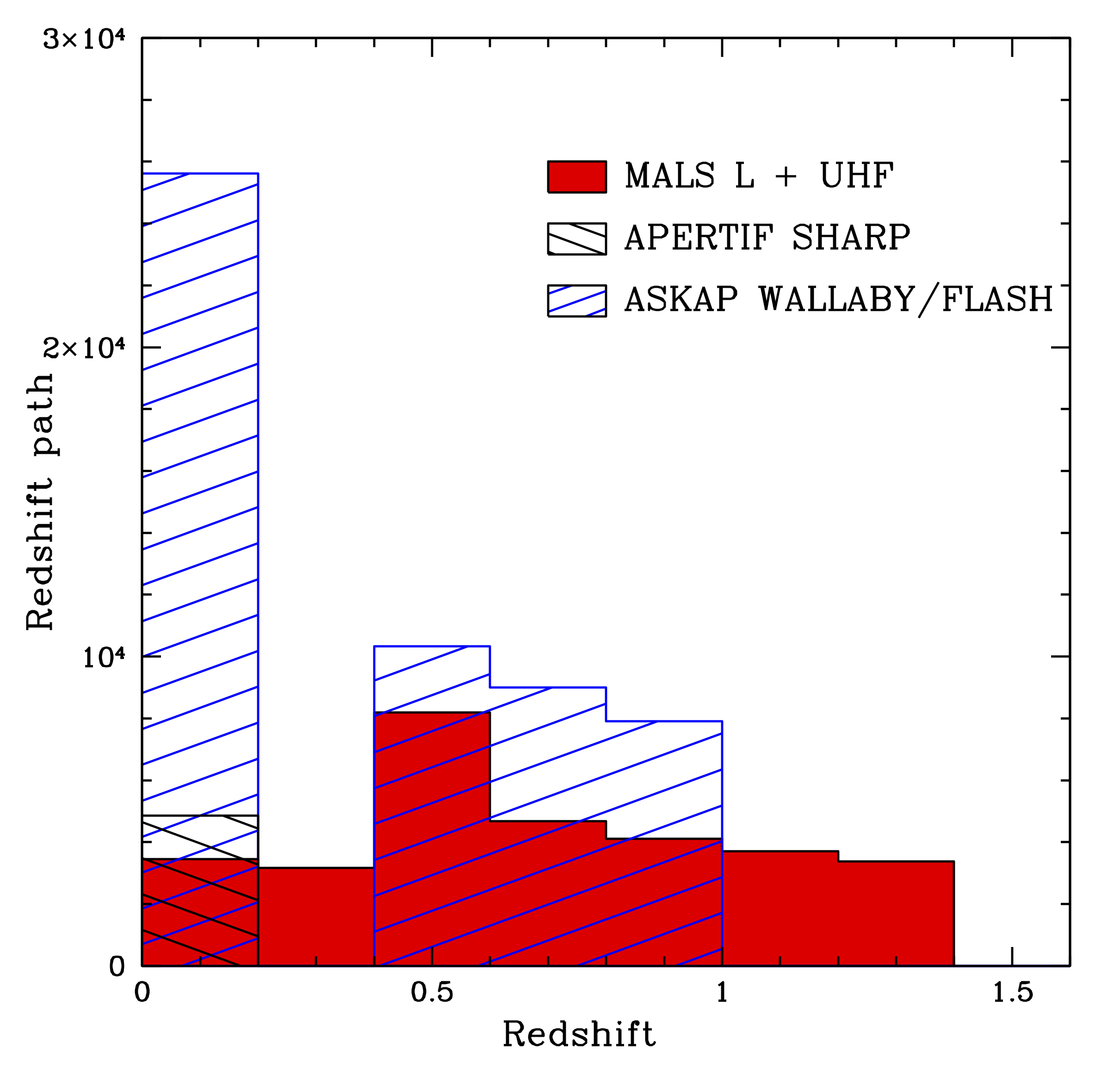}
\includegraphics[width=70mm, angle=0]{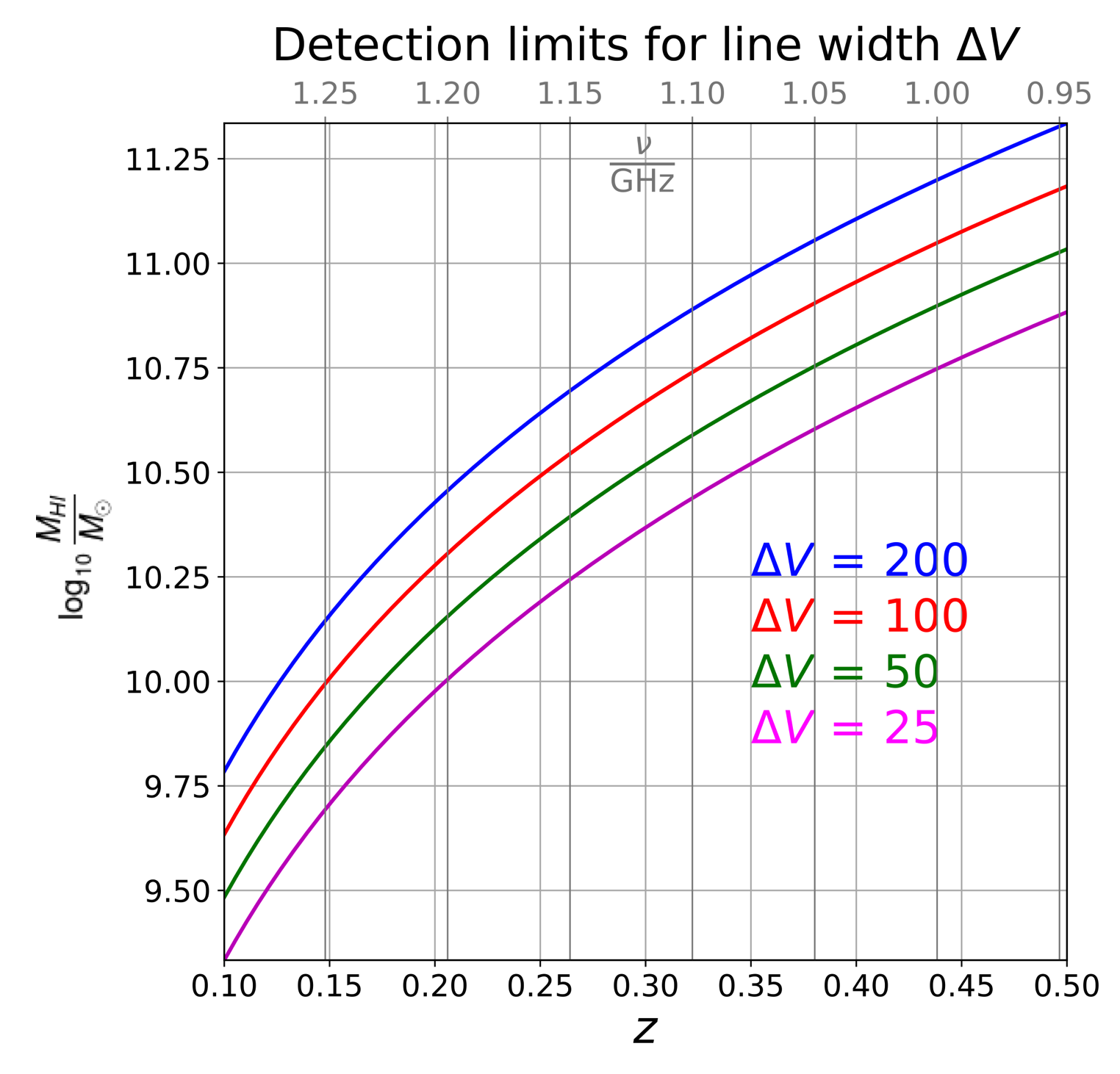}
}
\caption{
{\it Left:}
Redshift path comparison for various absorption line surveys (also see Table~\ref{abs}). The MALS L- and UHF-band 
$\Delta$$z$ are estimated for the primary beams at 1200\,MHz and 800\,MHz respectively. For MALS, the steep rise in the 
bin at $z=0.5$ is (primarily) due to this and the fact that radio sources are brighter at lower frequencies 
(spectral index$\sim$0.8). Also, at $z=0.5$ bin both the L- and UHF-bands contribute as they overlap at 900-1015\,MHz.
{\it Right:}
5$\sigma$ \hi\ mass emission-line detection limits for the observed fields (r = 0.36$^\circ$) for a total 
observing area of 300\,deg$^2$.
}
\label{dist}
\end{figure}
%%%%%%%%%%%%%%%%%%%%%%%%%%%%%
%%
%%

%%%%%%
The absorbers from MALS will be 
followed-up with (1) VLBI to map the pc-scale structure in the ISM \citep[][]{Srianand13dib}, 
(2) HST, SALT and VLT to detect various atomic/metal absorption lines to constrain the ambient radiation 
field, metallicity, density and temperature via detections, and VLA/MeerKAT to probe the magnetic fields in 
galactic discs via rotation measure synthesis \citep[][]{Farnes14}. These parameters, in addition to $n_{21}$ 
and $n_{OH}$, will be valuable inputs to test and refine the physical models of the ISM so far based primarily 
on the galactic ISM \citep[][]{Wolfire03} and understand the processes driving the evolution of SFR density. 
Further, the multi-wavelength follow-ups of the absorbers using ALMA, NOEMA, VLA, VLT and SALT will lead to 
the discovery of rare atomic and molecular species which will be used to measure the evolution of CMB temperature 
and constrain the effective equation of state of decaying dark energy \citep[][]{Noterdaeme10co}.

%%%%%%
\item{\bf Fuelling of AGN, AGN feedback and dust-obscured AGNs:} 
AGN activity is known to peak during the same epoch as SFR density. Both theoretical and observational 
considerations suggest a close relationship between galaxy evolution and AGN activity \citep[][]{Fabian12}. 
Further, the existence of red i.e. dust-obscured AGNs is predicted in both the orientation-based AGN unification 
models and the evolutionary scenarios but the exact fraction of such AGNs is still uncertain with the estimates 
ranging over 10-50\% \citep[][]{Glikman07}. Through a dust-unbiased search of \hi\ and OH absorptions, MALS 
will determine the (1) distribution and kinematics of atomic and molecular phases associated with circumnuclear gas 
and the ISM of AGN hosts, and (2) the fraction of dust obscured AGNs missed out in optical/ultraviolet surveys. 
Through these the survey will address several fundamental issues in the context of AGN evolution, AGN feedback and 
the AGN Unification paradigm.

Presently, about 100 associated \hi\ 21-cm absorbers and only a few OH absorbers are known \citep[][]{Baan85, Gupta06, Gereb15}. 
In MALS survey area, there will be $\sim$2000 sources brighter than 30\,mJy ($\sim$1 GHz) and suitable for the 
associated 21-cm and OH absorption line search. Adopting a detection rate of 25\% \citep[][]{Gereb15, Maccagni17}, the detection 
of $\sim$500 associated 21-cm absorbers is expected. This will provide a large sample to study the properties 
of cold gas in powerful AGNs ($>10^{24}$ W\,Hz$^{-1}$) by splitting these into subsamples of different AGN types 
and redshifts. For the brighter sources, the survey will also be sensitive to detect rare blue-shifted \hi\ 
wings in the absorption profile \citep[][]{Morganti05}. Such wings representing outflows have been identified 
in $\sim$5\% of the AGNs observed at $z<0.2$. The evolution of this fraction and the mass-outflow rate with 
redshift is a key observable to compare with the predictions from the models of galaxy evolution. MALS will 
extend the observations of these outflows to a largely unexplored ($z>0.2$) and the most relevant redshift 
range in this context.

%%%%%%
\item {\bf Galaxy evolution through \hi\ and OH emission:}
With an integration time of 56 minutes per pointing in L-band, MALS is a substantial  
blind emission line survey. MALS provides a set of 740 quasi-random pointings (with a potential sub-sample
pointing towards regions of high interest such as galaxy groups). With the natural rms of 0.26\,mJy\,beam$^{-1}$ 
over 20\,\kms\ at the pointing centre delivered by MALS, a galaxy like Leo T with a dark matter mass that is 
just enough to retain baryons, M(\hi)$\sim$2.8$\times$10$^5$\,M$_\odot$ and radius$\sim$240\,pc 
\citep[][]{Ryan-Weber08}, would be detectable above a 5$\sigma$ level and resolved to a distance of 4\,Mpc. 
Over an area of 300\,deg$^2$, MALS will be able to detect an $M_*$ galaxy up to a redshift of 
$z\sim$0.1-0.2 and the galaxies with the largest \hi\ masses out to $z > 0.5$ (see Fig.~1). We estimate 
the number of detections to be of the order of 5000-15000 galaxies, many of which will be resolved. In addition, 
tenuous gas at column densities of 0.2 M$_\odot$pc$^{-2}$ will be detectable over 20\,\kms\ at the 6$\sigma$ level 
at 45$^{\prime\prime}$ resolution. 
MALS will therefore probe the gas distributions and kinematics of a substantial number of galaxy disks while providing 
an unique opportunity to relate these to the cold atomic and molecular gas detected through \hi\ and OH absorption.

%%%%
The main goals of MALS as an \hi\ emission line survey are: (1) to catalog basic emission line parameters of observed 
objects, (2) to produce \hi\ emission line maps and data cubes of resolved emission line objects, (3) to parameterize 
the extended morphology and kinematics, spin characteristics, rotation curves, and surface density profiles of the 
resolved galaxies independently as well as in connection with the absorption line survey to understand the nature 
of 21-cm and OH absorbers, and (4) to study velocity and mass functions, (dark) matter distribution and quantify 
accretion properties. 
With these observables MALS will be a perfect emission line 
progenitor survey for already addressing major questions raised in the SKA science book: (1) studying the radial gas 
profiles of galaxies in the light of galaxy evolution \citep[][]{DeBlok15}, (2) studying the matter 
distribution in galaxies, (3) quantifying the intrinsic and mutual spin distribution in galaxies and its dependence 
on their environment \citep[][]{Obreschkow15}, and (4) studying the low column density environment of galaxies to 
quantify (AGN) feedback and accretion processes on a statistical basis \citep[][]{Popping15}.

%%%%
Besides \hi\ emission, MALS will also be capable of detecting OH Megamasers (MMs). Powerful OH MMs have already been 
detected up to a redshift of 0.265 (1318\,MHz; \citep[][]{Darling02}) and the emission profiles show strong outflows 
and evidences of strong interactions. The population of sources will increase at least up to a redshift of 2.5 because 
of the link between ULIRG\footnote{Ultraluminous infrared galaxy} host galaxies and the increasing merger rate with 
redshift. OH MMs are a good tracer of 
starburst activity and merger history and MALS will establish their luminosity function at higher luminosities and redshifts.

%%%%%%%
\item{\bf Radio continuum and polarization science with MALS:}
Magnetic fields are known to be present over a wide range of physical scales in astrophysical objects. 
They are an important component in the energy balance of the interstellar medium (ISM) of galaxies. 
Besides being important in determining the conditions for the onset of star formation in protostellar 
clouds they are also relevant for (1) the formation and stabilisation of gas disk and spiral arms, 
(2) the heating of gas, especially, in the outer disks and halos, and (3) the launching of galactic 
outflows and determining how the matter and energy are distributed throughout the galaxy. Overall, magnetic fields 
are tightly coupled to various stellar feedback processes and play an important role in a wide range of 
physical processes that drive galaxy evolution.  MALS will provide extremely sensitive wide-band L- and 
UHF-band observations to observe total intensity and polarized emission from a large number of normal and 
active galaxies enabling wide range of projects some of which are described below. 

{\bf Clusters, halos and relics:} MALS will allow us to conduct a blind survey of a statistically significant 
number of galaxy clusters to discover new diffuse cluster emission and constrain their formation models through 
spectral studies. We expect to detect about 770 clusters with M$_{500}$$>3\times10^{14}$\,M$_\odot$   
within the survey region, where M$_{500}$ is the cluster mass within a density five hundred times the critical 
density. All MALS pointings will be within the footprint of the {\it Advanced} ACT cosmic microwave background survey 
\citep[][]{Henderson16}, in which MALS team members are actively involved, and which will deliver Sunyaev-Zel'dovich 
masses for all clusters above this mass limit. With the $\sim$3\,$\mu$Jy flux density limit of MALS, we expect to observe diffuse 
radio halo emission in approximately 80 of these clusters, over a wide range of cluster mass and redshift 
($0<z<0.8$). This will provide a large enough statistical sample to study the evolution of radio halo 
properties, which significantly extends the existing heterogeneous sample of about 50 giant radio halos 
\citep[][]{Yuan15} beyond the current range of masses (M$_{500} > 5\times10^{14}$\,M$_\odot$) and redshifts 
($0.2 < z < 0.4$). We will place constraints on radio halo formation models, by measuring fluxes and spectral 
indices over the wide MeerKAT frequency bands, as well as distinguish between turbulent halos generated by 
re-acceleration during cluster mergers, and off-state halos of a purely hadronic origin found in more relaxed 
systems \citep[][]{Cassano12}. The polarization data from MALS will be crucial for a full study of the relic 
emission by testing the shock- related formation theories through inferred Mach numbers and magnetic field 
orientations.

{\bf Magnetic fields in galaxies and AGNs:} With the sensitivity of $\sim$3\,$\mu$Jy, we expect more than 30 polarized sources 
per square degree at 1.4 GHz \citep[][]{Rudnick14}. The wide-band coverage of MALS will allow us to quantify the 
effects of depolarization and measure currently unknown polarized source counts at low frequencies. The strong 
intervening 21-cm absorbers detected through MALS will be an excellent tracer of galactic discs and ideally 
suited to probe the evolution of magnetic fields in galaxies by rotation measure synthesis. Until now, such 
studies using samples of Mg~{\sc ii} absorbers have yielded ambiguous results \citep[][]{Farnes14} because 
of the small sample sizes and the difficulty in interpreting the origin of Mg~{\sc ii} absorbers.
Additionally, the sizes and structures of radio sources are often found to be affected by the ambient gas in the 
central regions of host galaxies, suggesting strong dynamical interaction with the external medium. Evidence of 
this gas may also be probed via polarization measurements and will be used to understand the role of jet-mode 
feedback in driving the outflows detected in 21-cm absorption and the evolution of AGNs.

%%%%%%
\item{\bf Constraining space- and time-variation of fundamental constants:} 
The absorption lines seen in the spectra of distant QSOs can be used to place constraints on the space and 
time variations of different dimensionless fundamental constants of physics \citep[][]{Uzan03}. For the past 
decade or so based on systematic efforts using high resolution echelle spectrographs at VLT and Keck, the 
community has constrained the variation of $\alpha$={\it (e$^2$/hc)} and $\mu$ = {\it (m$_e$/m$_p$)} 
at the level of 1 ppm at high-redshift. 
The next major step is to improve these constraints by another factor of 10 i.e. 1 part in $10^{7}$ 
which will be comparable to local measurements based on atomic clocks. The major systematics that prevents 
any further improvement stems from the stability of wavelength scales in present day optical spectrographs 
\citep[][]{Whitmore15}. Further progress is possible through radio absorption lines because the 
frequency scales at radio telescopes are known to be well-defined and compared to 
optical/ultraviolet lines the radio lines are more sensitive to the variation of fundamental constants. 
But a large sample of radio absorbers is required to achieve this so that any systematics introduced due to 
ionisation, radiation and excitation inhomogeneities in the gas that gets reflected as the variation of 
constants is randomised. MALS will provide such a sample for the first time. Through main- and 
satellite- OH absorption lines from the survey, or the joint analysis of 21-cm and OH absorption with 
various metal, atomic and molecular absorption lines detected through follow-ups with ALMA, NOEMA, VLA, 
VLT and SALT the survey will lead to tightest constraints on the variations of fundamental constants of physics.

%%%%%%
\item{\bf Stacking line, continuum and polarization data:}
Stacking of undetected \hi\ 21-cm line spectra of sources with known redshifts has become a tested method to 
study the atomic gas in galaxies \citep[][]{Lah07}. We will apply stacking technique to spectral line, radio 
continuum and polarization data from MALS to make measurements that would otherwise be possible only with 
much deeper observations or SKA-I. In \hi\ and OH absorption, MALS will likely detect galaxies and AGNs undetected 
at other wavelengths. This will not only constrain the redshift of such sources but also investigate their 
multi-wavelength properties by stacking.

%%%%
\end{enumerate}
%%%%

%%%%%%%%%%%%%%%%%%%%%%%%%%%%%%%%%%%%%%%%%%%%%%%%%%%%%%
\section{Survey Design}
\label{sec:design}
%%%%%%%%%%%%%%%%%%%%%%%%%%%%%%%%%%%%%%%%%%%%%%%%%%%%%%

%%%%%
%%%%%%%%%%%%%%%%%%%%%%%%%%%%%%%%
\begin{table}
\centering
%\begin{minipage}{120mm}
\caption{MALS Survey Design}
\begin{tabular}{ccccccc}
\hline
%Target & Date &  Exposure & Rot Sky& Air mass & Grism & Slit width   \\
{\large \strut}  MALS phase    & Number       &   Time        &  Spectral & Continuum & Total on-source \\
                               & of pointings & per pointing  & rms$^\dag$& rms       &     time   \\
                               &              &     (mins)    & (mJy\,beam$^{-1}$)& ($\mu$Jy\,beam$^{-1}$) &  (hrs) \\
\hline           
{\large \strut}  L-band                &  740 & 56  & 0.5  & 3 &  691 \\ 
                 (900-1670\,MHz)                                      \\ 
                 UHF-band              &  370 & 121 & 0.6  & 3 &  746 \\ 
                 (580-1015\,MHz)                                      \\ 
\hline
\end{tabular}
\flushleft{{
\small $^\dag$ Estimated at $\sim$1200\,MHz and $\sim$800\,MHz for the full band split into 32768 channels. 
}} 
\label{survey}
\end{table}
%%%%%%%%%%%%%%%%%%%%%%%%%%%%%%%%

The MALS survey design is summarized in Table~\ref{survey}.  The total L- and UHF-band on-source times 
for 740 and 370 pointings are 691 and 746\,hrs respectively.  Assuming 15\% overheads this corresponds to a total observing time 
of 1655\,hrs.  
The survey strategy that yields almost uniform redshift path coverage over the entire redshift range of interest 
(see Fig.~1) is based on the optimization of the following two parameters that drive the number of detections and the 
science output from the survey:
%%%
\begin{enumerate}
%%%
\item {Optical depth sensitivity ($\int\tau$dv):} The target 5$\sigma$ integrated optical depth sensitivity 
for intervening \hi\ 21-cm absorption at the center of the primary beam is $\int\tau$dv = 0.045\,\kms. This 
corresponds to a sensitivity to detect the CNM (100\,K) in N(\hi)$\sim$10$^{19}$\,cm$^{-2}$ gas via \hi\ 21-cm absorption, 
and is essential to detect the CNM in a wide range of physical environments.  This will be achieved by centering 
L- and UHF-band pointings on bright ($>$400\,mJy at 1\,GHz) radio sources.    

\item {Redshift coverage/ Survey redshift path ($\Delta$$z$):} 
As the sensitivity to detect the CNM down to a \hi\ column density of 10$^{19}$\,cm$^{-2}$ will be 
achieved only towards the central source, it is essential that these be selected carefully. 
The MeerKAT L- and UHF-band receivers cover frequency ranges of 900-1670\,MHz and 580-1015\,MHz respectively.
For the \hi\ 21-cm line, these correspond to a redshift coverage of $0<z<0.58$ and $0.40<z<1.44$, 
respectively\footnote{The corresponding redshift ranges for OH main lines are $0<z<0.85$ 
and $0.64<z<1.87$, respectively.}.  
The redshift path for central bright sources will be maximized by selecting these to be at $z>0.6$ and 
$z>1.4$, respectively. With this approach, based on the $n_{21}$ from \citep[][]{Gupta12}, which has $\sim$50\% errors, 
we expect to detect $\sim$100 intervening 21-cm absorbers towards the central bright sources (we adopt $n_{21}$=0.13; 
$\Delta$z = 800 and assume no redshift evolution).

%%%
The spectral rms of $\sim$0.5\,mJy would also ensure that an adequate redshift path to deliver a similar number of 
intervening 21-cm absorbers with N(\hi)$\sim$10$^{20}$\,cm$^{-2}$ (i.e. $\int\tau$dv=0.5\,\kms\ for 100\,K) 
is achieved towards off-axis weaker sources, and the sight lines with rare high column 
density absorbers are not under-represented in the survey.
%%%
\end{enumerate}
%%%

%%%%
Over the last few years results from several \hi\ 21-cm absorption line searches in the  samples 
of Mg~{\sc ii} absorbers at $0.5<z<1.5$ \citep[][]{Kanekar09mg2, Gupta12}, DLAs at $z>2$ 
\citep[][]{Curran10, Srianand12dla, Kanekar14} 
and quasar-galaxy pairs (QGPs) at $z<0.3$ \citep[][]{Gupta10, Borthakur10, Reeves16, Dutta16} have become available. 
From VLBA imaging and spectroscopy of a subset of these samples it has become increasingly obvious that 
higher detection rates of intervening 21-cm absorption are achieved for flat-spectrum radio sources 
(see Fig.~6 of \citep[][]{Gupta12}). This is related to the typical correlation length of 30-100\,pc 
exhibited by the CNM gas (see also \citep[][]{Braun12}). The gas clouds are expected to be much more 
compact ($<$10\,pc) in the case of translucent or dense phases of the ISM \citep[][]{Srianand13dib}. 
Therefore, for a blind search of intervening absorbers the background sources have to be as compact 
as possible. The flat-spectrum radio quasars (FSRQs), which are known to be compact, are hence the straightforward choice for MALS.

Thus, in addition to the criteria that the central sources for L- and UHF-band pointings be at 
$z>0.6$ and $z>1.4$, respectively, we further require that a majority of these be FSRQs.

%%%%%%%%%%%%%%%%%%%%%%%%%%%%%%%%%%%%%%%%%%%%%%%%%%%%%%
\section{Survey Strategy, Observing Plan and Data Releases}
\label{sec:data}
%%%%%%%%%%%%%%%%%%%%%%%%%%%%%%%%%%%%%%%%%%%%%%%%%%%%%%

The survey is expected to start in 2018 and will be carried out over a period of five years.  Based on the outcomes 
from the first year of observations, the proportion of L- and UHF-band pointings for the survey will be revised. 
%(see Table~\ref{obsplan} for the Observing Plan).

%%%%%%%
\vspace{0.2cm}
{\bf Survey strategy and observing setup}
%%%%
\begin{enumerate}
%%%
\item {\bf Observing setup:} The data will be acquired in both parallel- and cross-polarization hands using a
correlator dump time of 4\,s. The L (900 - 1670\,MHz) and UHF (580 - 1015\,MHz) bands will be split into 32768 
frequency channels.  This would deliver a spectral resolution of $\sim$5\,km\,s$^{-1}$, and will be 
adequate to detect and resolve the structure in intervening 21-cm absorption lines (e.g. see Fig. 
12 of \citep[][]{Gupta09}).

%%%
\item {\bf Calibration overheads:} For a two-hour observing block we assume a calibration overhead of
15\%. This includes observations of a standard calibrator every two hours and a 2\,min scan on a phase 
calibrator every 45\,mins for flux density scale and bandpass calibrations. 
The leakage calibration will be carried out through observations of an unpolarized calibrator.  We will also 
require a polarization angle calibrator. It is quite possible that the polarization angle can be calibrated 
using data from previous days, and that there will be no need to observe a separate phase calibrator.
All this will be tested and confirmed during the commissioning phase.

%%%
\item {\bf Observing blocks and Scheduling:} The time spent on each pointing ranges from 56 to 121 mins. 
The total time on each pointing may be split into two equal sessions, which will be executed 
within a few months with a gap that ensures a shift of 20-30\,\kms\ in the absorption lines. This shift due to the 
heliocentric motion of the Earth would allow us to unambiguously distinguish between true absorptions and RFI, and 
is required to build a reliable absorption line catalog. We intend to execute these sessions through observing blocks 
of one to several hours.

%%%
\item {\bf Cadence:} Each pointing will have two epochs separated by a few months. The pointings ($\sim$300) that are 
common between the L- and UHF-band phases will have four epoch observations.   

%%%%
\item {\bf Strategy for the selection of pointings:} For the first year of observations, we will
select ~70\% of the pointings at $-40^\circ  < \delta < +30^\circ$ and 100\% of these will be centered at FSRQs with 
spectroscopically confirmed redshifts. This will allow accurate determination of redshift path length functions 
towards the central bright source and lead to the first reliable constraints on $n_{21}$($z$) and $n_{OH}$($z$). 
The availability of VLBI images for all of the central targets in the first year of observations will allow us to quantify the effect of 
radio structure on the detectability of absorption and its redshift evolution. All this will allow us to use 
first year's observations as a proving ground and, if needed, alter the choice of pointings for the 
subsequent phases.

%%
%%
%%%%%%%%%%%%%%%%%%%%%%%%%%%%
\begin{figure}
%\hbox{
\begin{center}
\includegraphics[width=120mm, angle=0]{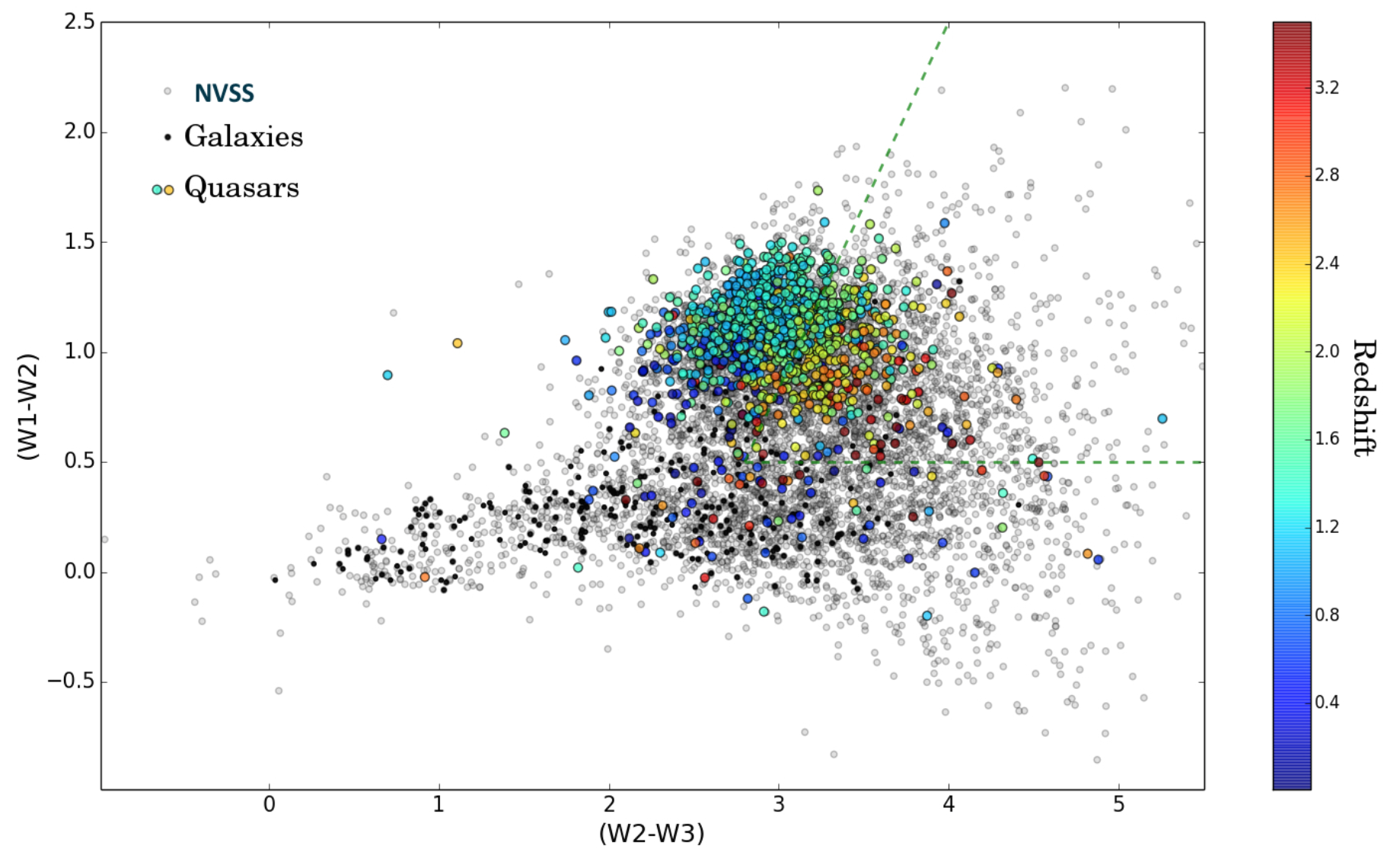}
%\includegraphics[width=95mm, angle=0]{hiem.png}
%}
\end{center}
\caption{
%{\it Left:}
WISE IR colors for radio sources ($\delta<+20^\circ$, $>$200\,mJy) (grey). Galaxies identified in the literature are 
shown as small black points and the known quasars are color coded as a function of redshift. The dashed lines indicate 
the color-color space for selection of $z>1.4$ radio loud quasars.
}
\label{wise}
\end{figure}
%%%%%%%%%%%%%%%%%%%%%%%%%%%%%
%%
%%

%%%%
Bright FSRQs with known spectroscopic redshifts, especially in the southern hemisphere, are rare.    
Therefore, we are carrying out an ambitious optical spectroscopic campaign using SALT and NOT 
to increase the number of FSRQs with $z>1.4$.
For this, using WISE IR colors we have developed a pre-selection method that allows us to select $z>1.4$ 
`candidate' quasars with $\sim$75\% efficiency (Fig.~\ref{wise}). 
Through this ongoing program we have already identified $\sim$100 new FSRQs over last two years. 
%%%

\end{enumerate}
%%%%%%

All the data products from the survey (i.e. calibrated visibilities and images) will be released to community 
through regular data releases.

%%%%%%%%%%%%%%%%%%%%%%%%%%%%%%%%%%%%%%%%%%%%%%%%%%%%%%
\section{Relationship with other SKA Pathfinder surveys}
\label{sec:ska}
%%%%%%%%%%%%%%%%%%%%%%%%%%%%%%%%%%%%%%%%%%%%%%%%%%%%%%

Understanding galaxy evolution through \hi\ 21-cm line is one of the key science drivers of 
the SKA \citep[][]{Staveley15, Morganti15}. 
MALS will measure the evolution of the cold atomic and molecular gas cross-sections of galaxies in a 
luminosity- and dust-unbiased way and constrain the ISM physics at pc scales over the entire redshift 
range: $0<z<2$ (cf. Table~\ref{abs} for complementarity with other 21-cm absorption line surveys).
Through the excellent sensitivity of MeerKAT over a wide range of spatial resolutions and wideband 
spectro-polarimetry, MALS will simultaneously deliver a highly competitive blind \hi/ OH emission 
line and radio continuum survey, and will address a wide range of science issues raised in the SKA science book.   
The survey will also complement \hi\ 21-cm emission line surveys such as 
(1) LADUMA and WALLABY that will be affected by luminosity bias 
and will probe only most luminous objects at higher redshifts, and (2) MHONGOOSE that will 
measure the sub-kpc scale properties of gas but in a sample of nearby galaxies. The conversion of gas into 
stars involves complex physical processes from pc to kpc and even Mpc scales. Together, these surveys will  
provide observational constraints on the properties of cold gas and the underlying physical processes at all 
physical scales required to understand the evolution of SFR density and design the future \hi\ 21-cm line surveys with SKA.

%%
%%%%%%%%%%%%%%%%%%%%%%%%%%%%%%%%%%%%%%%%%%%%%%%%%%%%%%
\acknowledgments{\noindent 
We thank Kenda Knowles of UKZN for useful discussions.
We thank IUCAA, Rutgers University and the South African members for the SALT observing time 
contributed to the survey to identify high-$z$ quasars for MALS.
NOT is operated by the Nordic Optical Telescope Scientific Association at the Observatorio 
del Roque de los Muchachos, La Palma, Spain, of the Instituto de Astrofisica de Canarias.
We acknowledge the support from ThoughtWorks India Pvt. Limited in developing ARTIP - a prototype of MALS data 
analysis pipeline.
}

%%%%%%%%%%%%%%%%%%%%%%%%%%%%%%%%%%%%%%%%%%%%%%%%%%%%%%

%%%%%%%%%%%%%%%%%%%%%%%%%%%%%%%%%%%%%%%%%%%%%%%%%%%%%%
%\begin{thebibliography}{99}
\bibliographystyle{JHEP}
\bibliography{/Users/ngupta/Desktop/Comp/mybib.bib}
%\end{thebibliography}

%%%%%%%%%%%%%%%%%%%%%%%%%%%%%%%%%%%%%%%%%%%%%%%%%%%%%%
\appendix
\section{Author Affiliations}
\noindent
{\small {\it 
$^1$ Inter-University Centre for Astronomy and Astrophysics, India; 
$^2$ ASTRON, the Netherlands Institute for Radio Astronomy, The Netherlands;  
$^3$ Rutgers, the State University of New Jersey,  USA;
$^4$ University of Manchester, UK;
$^5$ National Radio Astronomy Observatory, USA;
$^6$ Laboratoire d'Astrophysique de Marseille, France; 
$^7$ Physics and Astronomy Department, University of Western Cape, South Africa; 
$^8$ Observatoire de Paris, France;   
$^9$ Centre for High Performance Computing, South Africa; 
$^{10}$ Dark Cosmology Center, Niels Bohr Institute, Denmark; 
$^{11}$ CSIRO Astronomy and Space Science, Australia;  
$^{12}$ Astrophysics and Cosmology Research Unit, University of KwaZulu Natal, South Africa; 
$^{13}$ Astrophysics, University of Oxford, UK; 
$^{14}$ Square Kilometre Array South Africa, South Africa; 
$^{15}$ National Centre for Radio Astrophysics, India; 
$^{16}$ Argelander-Institut f\"{u}r Astronomie (AIfA), Universit\"{a}t Bonn, Germany; 
$^{17}$ Max-Planck-Institut f\"{u}r Radioastronomie, Germany; 
$^{18}$ Institut d'Astrophysique de Paris, France;  
$^{19}$ University of South Carolina, Department of Physics and Astronomy, USA; 
$^{20}$ European Southern Observatory, Chile; 
$^{21}$ Astronomy Department, California Institute of Technology, USA;  
$^{22}$ South African Astronomical Observatory, South Africa; 
$^{23}$ INAF-Osservatorio Astronomico di Cagliari, Italy; 
$^{24}$ Department of Physics, Royal Military College of Canada, Canada; 
$^{25}$ Kapteyn Astronomical Institute, University of Groningen, The Netherlands; 
$^{26}$ University of Utah, USA; 
$^{27}$ International Centre for Radio Astronomy Research, The University of Western Australia, Australia.
}}
%%%%%%%%%%%%%%%%%%%%%%%%%%%%%%%%%%%%%%%%%%%%%%%%%%%%%%
\end{document}